\documentstyle[preprint,version2,aps]{revtex}

\begin{document}
\draft
\begin{title}
 Universal Diamagnetism  of Charged Scalar Fields
\end{title}
\author{ Debnarayan Jana \cite{Auth}}
\begin{instit}
Raman Research Institute, Bangalore 560 080, India
\end{instit}

\begin{abstract}
We show  that charged scalar fields are always diamagnetic, even in the
presence of interactions and at finite temperatures. This generalises
earlier work on the diamagnetism of charged spinless bosons to the case of
infinite degrees of freedom.
\end{abstract}

\newpage
\section{Introduction}
A classical gas of charged point particles is non-magnetic, by Van
Leeuwen's Theorem~\cite{van}. But quantum mechanically, free spinless Bose
particles in a uniform magnetic field show diamagnetism\cite{lan}. One
finds that  the energy of the system in a magnetic field is higher than
in the absence of the field. Simon\cite{sim} proved quite generally that
the ground state  of  non-relativistic spinless Bosons interacting through
an arbitrary potential {\it always increases} in a  magnetic field. He
went on to extend this result  by showing that the free energy in the
presence of a magnetic field is always greater than the free energy in the
absence of a magnetic field at all temperatures~\cite{sim}. An alternative
proof of this result is given in ~\cite{ss}. All this  work  which deals
with systems with a finite number of degrees of freedom suggests that
diamagnetism is a universal property of Spinless Bosons.  In field theory
(which describes systems with an infinite number of degrees of freedom)
charged spinless Bosons are described by complex scalar fields. One might
therefore expect that charged scalar fields would also show diamagnetic
behaviour.  With this motivation we study the magnetic behavior of scalar
field theories.

 The paper is organised as follows. The first part deals with finite
temperature free  scalar field theory in the presence of an external
homogeneous magnetic field.  Here we explicitly calculate the partition
function and the free energy as a function of the applied magnetic field.
This expression is formally divergent. Using a suitable regularization
scheme we compute the {\it difference} in the free energy (with and
without the magnetic field) and obtain a finite answer. This difference is
also shown to be positive, thus establishing the diamagnetic behaviour of
free charged scalar fields.

We then move on to interacting scalar field theory in the second part.
Here, we cannot evaluate the partition function explicitly. Nevertheless
we prove the universal diamagnetism of scalar fields by assuming a finite
momentum cutoff in the theory. If the theory is renormalizable, then one
can take this cutoff to infinity while maintaining finiteness of all
physical quantities. In both  cases the results obtained are exact.

\section{Free case}
\par
In this section we  calculate the free energy of free scalar fields
in the presence of an external uniform  magnetic field. For ease of
presentation we work in two spatial dimensions. The interesting physics
takes place in the plane  normal to the applied  field. Generalization to
higher dimensions is straightforward.

Let $\Phi$ be a complex scalar field which describes charged spinless Bosons.
 The Lagrangian density of a free charged scalar field in the
 presence of a constant homogeneous external magnetic field is given by

  \begin{equation}
 {\cal L}=(D_{\mu}\Phi)^\ast(D^\mu\Phi) - m^2(\Phi^\ast\Phi)
 \end{equation}
    where $\mu=0,1,2$,
 \begin{equation}
  D_{\mu} = \partial_{\mu}-{\rm i} e A_{\mu}
 \end{equation}
 and $m$ and $e$ are the mass and charge respectively.
 (We set $\hbar=1$ and $c = 1$).
  Now, we write the complex field in term of two real fields $\Phi_{1}$
 and $\Phi_{2}$.
 \begin{equation}
 \Phi=\frac{\Phi_{1}+i\Phi_{2}}{\sqrt 2}\nonumber,
 \Phi^\ast=\frac{\Phi_{1}-i\Phi_{2}}{\sqrt 2}
 \end{equation}
 This theory has a global U(1) symmetry and therefore a conserved Noether
 charge ${\rm Q}$,
 given by
 \begin{equation}
  Q=\int d\,^2x (\Pi_{1}\Phi_{2}-\Phi_{1}\Pi_{2})
  \end{equation}
   where
   \begin{equation}
    \Pi_{i}=\partial_{0}\Phi_{i}
   \end{equation}

 The Hamiltonian density of the system is given by
 \begin{eqnarray}
  {\cal H}
&=&\frac{1}{2}(\Pi_{1}^2+\Pi_{2}^2)+\frac{1}{2}(\nabla\Phi_{1})^2+\frac{1}{2}(\nabla\Phi_{2})^2+\nonumber\\
           & &~~~~~~\frac{1}{2}(m^2+e^2A^2)(\Phi_{1}^2+\Phi_{2}^2)-j\cdot A,
 \end{eqnarray}
 where the current density j is given by
 \begin{equation}
 {\bf j}= {\rm -e}(\Phi_{1}{\bf\nabla}\Phi_{2}-\Phi_{2}{\bf\nabla}\Phi_{1}).
 \end{equation}
 We now suppose that the external magnetic field is uniform in the $x-y$ plane.

We choose the temporal gauge ( $A_{0}=0$).
   The constant magnetic field $B$ is
     \begin{equation}
    B = \partial_{x}~A_{y}- \partial_{y}~A_{x}
   \end{equation}
 where ${\bf A(x)}$ is independent of t.

  The action of this theory is

\begin{equation}
 S =\int_{0}^{\beta} \int d\,^2x d\tau  {\cal L}(\Phi,\Phi^\ast,A),
 \end{equation}
where  $ \tau$ is the imaginary time variable which runs
  from 0 to $\beta$ (=1/($k_{B}T$)), the inverse temperature.
 The action defined above is quadratic and so the partition function can
be evaluated exactly.
 As is usual in finite temperature field theory~\cite{ftft}, we impose
 periodic boundary conditions for  Bosonic fields
 \begin{equation}
      \Phi({\bf x},0) = \Phi({\bf x},\beta).
 \end{equation}

 Now, the partition function of this theory can be written as
 \begin{eqnarray}
 Z(B)&=&\int {\cal D}[\Pi_{1}] {\cal D}[\Pi_{2}]\int {\cal D}[\Phi_{1}]
{\cal D}[\Phi_{2}] \exp\left[ \int d\tau d^2x \right. \nonumber \\
& &\left.\left({\rm i}\Pi_{1}\frac{\partial\Phi_{1}}{\partial\tau}+  {\rm
i}\Pi_{2}\frac{\partial\Phi_{2}}{\partial\tau}
-H(\Phi_{1}, \Phi_{2}, \Pi_{1}, \Pi_{2})+
\mu(\Phi_{2}\Pi_{1}-\Phi_{1}
\Pi_{2}) \right) \right]
 \end{eqnarray}
 Here $\mu$ is the chemical potential associated with the conserved
charge.  We pick the gauge in which the vector potential {\bf A} is (-By, 0),
and expand the complex scalar field in terms of modes adapted to the
present situation.  These modes solve the  Klein-Gordon equation in an
external magnetic field. The eigenfunctions are labelled by one discrete
($l$) and one continuous $p_{x}$ quantum number and the spectrum depends
on $l$ only.  In the gauge we choose, the modes are plane waves  in the x
direction and harmonic oscillator (i.e. gaussian) wavefunctions in the y
direction.

The spectrum is given by
 \begin{equation}
  \omega_{l}^2 =  m^2 + (2l+1)eB,~~~ l=0,1,2.......\infty
  \end{equation}
The degeneracy of these states for fixed $l$ is ${e A B}/{2\pi}$ ,
where $A$ is the area of the system.  So, these modes can be thought of
as quantized harmonic oscillators. Expanding
the fields $\Phi_{1}$ and $\Phi_{2}$ in these modes the system reduces to
a collection of harmonic oscillators with frequency $\omega_{l}$.

By standard manipulations ~\cite{ftft} we get the free energy as

\begin{eqnarray}
F(B)=-\ln{\rm Z(B)}/\beta&=&{\rm 2\pi eA}B\sum_{l=0}^\infty\left[\omega_{l} +
\frac{1}{\beta}\ln(1-{\rm exp}(-\beta(\omega_{l}-\mu)) \right.\nonumber\\
& &~~~~~~~~~~~~~~\left.+\frac{1}{\beta} \ln(1-{\rm
exp}(-\beta(\omega_{l}+\mu))\right]
 \label{sq}
 \end{eqnarray}

The first term in the square brackets corresponds to the zero point
fluctuation  of the vacuum and the other two terms are  finite
temperature contributions.
We will first compare the free energy of the system with and without the
magnetic field at zero temperature.

The free energy of the system in presence of the magnetic field at zero
temperature is given by
\begin{equation}
F_{0}(B)=2\pi A~eB\sum_{l=0}^\infty~\omega_{l},\label{cx}
\end{equation}
where $\omega_{l}^2=m^2+(2l+1)eB$. Obviously, this sum diverges. In order to
obtain a finite answer we need to impose a cutoff $L$ in the sum
(\ref{cx}). Then the free energy becomes
\begin{equation}
F_{0}(B,L)=2\pi A~eB\sum_{l=0}^L~\omega_{l}.\label{tre}
\end{equation}
The free energy in the absence of the magnetic field is also given by a
divergent expression
\begin{equation}
F_{0}(0)= 2\pi A\int_{0}^\infty~pdp~\sqrt{(p_{x}^2+p_{y}^2)+m^2}\label{vx}
\end{equation}
We regularize this expression by imposing a cutoff $\Lambda$. Thus the free
energy (\ref{vx}) becomes
\begin{equation}
F_{0}(0,\Lambda)=2\pi
A\int_{0}^\Lambda~pdp~\sqrt{(p_{x}^2+p_{y}^2)+m^2}\label{frt}
\end{equation}
In order to compare the free energies in equations (\ref{tre}) and
(\ref{frt}), we choose the cutoffs $L$ and $\Lambda$ in such a way that
both systems have the same number of modes. Such a procedure can be
justified on physical grounds if one imagines turning on the magnetic field
adiabatically.

Counting the modes upto the L-th Landau level we find
\begin{equation}
2\pi A~eB\sum_{l=0}^L~~{\rm 1} = 2\pi A~eB(L+1)
\end{equation}
Similarly, for the momentum cutoff upto $\Lambda$ we get the modes without
the magnetic field as
\begin{equation}
2\pi A\int_{0}^\Lambda pdp = \pi~A\Lambda^2\label{op}
\end{equation}
Equating these gives us
\begin{equation}
\Lambda^2=2~eB~(L+1)
\end{equation}
Now, the free energy in absence of the magnetic field depends on magnetic
field through the momentum cutoff and is given by
\begin{equation}
F_{0}(0,L)=2\pi A\int_{0}^{\Lambda(B)}~pdp~\sqrt{p^2+m^2}
\end{equation}
The difference between the two free energies is given by
\begin{equation}
\Delta F(B,L)=F_{0}(B,L)-F_{0}(0,L)
\end{equation}
We define
$G(B)={F_{0}(B,L)}/{2\pi A}$, $H(B)={F_{0}(0,L)}/{2\pi A}$ and
$K(B)=G(B)-H(B)$. Numerically evaluating these sums for large
but finite $L$ and plotting them in figure 1  and figure 2 shows
that $K(B)$ is the difference between two large quantities.
As the cutoff $L$ goes to infinity then $K(B)$ becomes the
difference between two infinities. In this limit we find that $K(B)$ tends
to a finite value.  Thus, the susceptibility at zero temperature in the
relativistic case is non-zero. This vacuum susceptibility can be
interpreted as due to virtual currents.

We now show analytically that $\Delta F(B)$ is positive i.e. the vacuum is
diamagnetic.  Note that
\begin{equation}
\Delta F(B)~=~F(B)-F(0)=\sum_{l=0}^\infty~a_{l}(B,m) \label{ok}
\end{equation}
where $a_{l}(B,m)$ is given by
\begin{equation}
a_{l}(B,m)=eB\left[\sqrt{m^2+(2l+1)eB}-\int_{0}^{1}~d\alpha~\sqrt{m^2+2(l+\alpha)eB}\right]
\end{equation}
Introducing a dimensionless quantity $\rho=\frac{eB}{m^2}$ the above
equation becomes
\begin{equation}
a_{l}(\rho)=\rho\left[\sqrt{1+(2l+1)\rho}-\int_{0}^{1}~d\alpha~\sqrt{1+2(l+\alpha)\rho}\right]
\end{equation}

The positivity of $a_{l}(\rho)$ for each $l$ can be proved geometrically.
Defining $z_{l}={(1+2l\rho)}/{2\rho}$ and $f(\alpha)=\sqrt{z_{l}+\alpha}$, the
coefficient  $a_{l}(\rho)$ can be rewritten in terms of $c_{l}(\rho)$ as
\begin{equation}
c_{l}(\rho)=\frac{a_{l}(\rho)}{\sqrt{2}~\rho^{3/2}}=f(1/2)-\int_{0}^1~d\alpha~f(\alpha).
\end{equation}
Since, the function $f(\alpha)$ is convex, the area under the tangent
drawn at $\alpha=1/2$ is greater than the area under the curve (see figure
3). This shows that $c_{l}(\rho)$ is positive. To show the
convergence of the sum (\ref{ok}) we note that
\begin{equation}
\int_{0}^1~d\alpha~f(\alpha)~\leq~\left[f(1/2)~-~\frac{f(0)+f(1)}{2}\right]
\end{equation}
Now, applying mean value theorem twice one can easily show that

\begin{equation}
\int_{0}^1~d\alpha~f(\alpha)\leq~-\frac{1}{16(z_{l}+\alpha)}{3/2}
\end{equation}
Thus the coefficient $c_{l}(\rho)$ is positive for each
$l$ and the sum converges, hence the diamagnetic inequality is established.

\par
Massless Limit:
The magnetisation at zero temperature in the zero mass limit is given
by
\begin{equation}
{\rm M(B)}\sim~-\sqrt{B}
\end{equation}
So, the susceptibility in this zero mass limit is given by
\begin{equation}
\chi(B)~\sim~-~\frac{1}{\sqrt{B}}
\end{equation}
which diverges~\cite{div,div1} as $B$ goes to zero. This feature of the
susceptibility has already been noticed in the magnetised pair Bose
gas~\cite{mag}.  So, in this case the external field will be totally
expelled. This happens because of large number of virtual particle and
antiparticle produced in the ground state so that the overall
diamagnetism of the system is high enough to totally expel the external
field.

\par
 Another interesting point  is that
there exists a critical magnetic field below which magnetic field will be
totally expelled. This critical field can be estimated as follows. The
effective magnetic field can be defined as $B_{eff}=~B + 2\pi~M$ Here, the
magnetisation $M$ varies as $-~const\sqrt{B}$.  Therefore, there exists a
critical magnetic field $B_{c}$ where the effective field $B_{eff}$
vanishes.
\par
Now, for the finite temperature case  one can regulate the
free energy through the same mode matching regularisation method. Finally,
 one can write down the free energy difference in dimensionless form as
before
\begin{equation}
\Delta F(B)= F(B)-F(0)=\sum_{l=0}^\infty~b_{l}(\rho,\delta,\zeta)
\end{equation}
where,
\begin{equation}
b_{l}(\rho,\delta,\zeta)=\frac{\rho}{\delta}\left[g(\rho,l,1/2)-\int_{0}^{1}~d\alpha~g(\rho,l,\alpha)\right].
\end{equation}
The dimensionless variables are defined as
$\delta=\beta m$ and $\zeta=\beta\mu$.

The coefficient $g(\rho,l,\alpha)$ is given by
\begin{eqnarray}
g(\rho,l,\alpha)&=&log\left(1-exp(-\delta(\sqrt{1+2(l+\alpha)\rho}-\zeta))\right)+\nonumber\\
              &
&log\left(1-exp(-\delta(\sqrt{1+2(l+\alpha)\rho}+\zeta))\right)\label{gb}.
\end{eqnarray}
Now, defining $z_{l}=\frac{\delta^2(1+2l\rho)}{2\rho}$ we can write the
equation (\ref{gb})
\begin{eqnarray}
g(\rho,l,\alpha)&=&log\left(1-exp(-(\sqrt{z_{l}+\alpha}-\zeta))\right)+\nonumber\\
              & &log\left(1-exp(-(\sqrt{z_{l}+\alpha}+\zeta))\right).
\end{eqnarray}
The function $g(\rho,l,\alpha)$ is convex and so the zero temperature
argument applies unchanged. It follows that
the free energy satisfies the following inequality
\begin{equation}
F(B) \geq F(0)
 \end{equation}
 Thus the  response of the system to the magnetic
  field will be diamagnetic.

 \section{Interacting case}
In this section  we want to extend the diamagnetic inequality to the
self-interacting field theory case including the dynamical interaction
between scalar fields. The partition function of this charged
self-interacting field theory in the presence of the magnetic field can be
written as
\begin{equation}
{\rm Z(B)}= \int\int{\cal D}[\Phi]{\cal D}[\Phi^\ast]{\rm
exp}(-S(\Phi,\Phi^\ast,A)), \label{fq}
\end{equation}
where  the action S is defined as
\begin{equation} {\rm S}= \int\int
d\,^2x~d\tau\left[(D_\mu\Phi)(D^\mu\Phi)^\ast +m^2(\Phi^\ast\Phi) +
V(\Phi,\Phi^\ast)\right].
\end{equation}
\par
 The action is not quadratic and $Z(B)$ cannot be evaluated in closed
form. Nevertheless, we show that the response of the system to an external
magnetic field is diamagnetic.  Since the formal expression for the
partition function may not exist (the integrals may not exist) we impose a
cut off in momentum space. The functional integral in (\ref {fq}) signifies
that one only integrates over those field configurations whose Fourier
transforms have support within a sphere of radius $\Lambda$ in momentum
space. The partition function then explicitly depends on $\Lambda$. We do
not explicitly indicate the $\Lambda$ and $\mu$ dependence of
$Z(B,\lambda,\mu)$ below.

We divide the action into two parts $S_{0}$ and $S_{int}$,
   where $S_{0}$ is the action in the absence of the external field.
    \begin{equation}
  {\rm S}= {\rm S_{0}} + {\rm S_{int}},
  \end{equation}
  where
  \begin{equation}
   {\rm S_{0}}=\int\int d\,^2x
d\tau\left[(\partial_\mu\Phi)(\partial^\mu\Phi)^\ast + m^2(\Phi^\ast\Phi)
                  +V(\Phi^\ast,\Phi)\right]\\,
  \end{equation}
  \begin{equation}
  {\rm S_{int}}= \int\int d\,^2x d\tau\left[{-\rm i} e
(\partial_\mu\Phi)(A^\mu\Phi^\ast)
                  +{\rm i}e(A_\mu\Phi)(\partial^\mu\Phi^\ast)+e^2({\bf
A\cdot\bf A})(\Phi\Phi^\ast)\right].
  \end{equation}
 Notice that $exp(-S_{0})$ is a positive measure on the space of field
 configurations. The ratio $Z(B)/Z(0)$ can therefore be regarded as
 the expectation  value of $exp(-S_{int})$. Since $exp(-S_{int})$ is
  an oscillatory function whose modulus is less than or equal to $1$, we
  conclude that
  \begin{equation}
\frac{\rm Z(B)}{\rm Z(0)}  =   \ll {\rm exp} {-S_{int}} \gg
                                \leq 1
\end{equation}

 This implies that
\begin{equation}
F(B)\geq F(0)
\end{equation}

In this derivation, we have not assumed any form for the vector potential.
So, the result derived above is true for both homogeneous or inhomogeneous
magnetic fields of any strength. Since $\beta$ is arbitrary, the result
holds at all temperatures. The argument presented here works for any
arbitrary interaction $V(\Phi^\ast\Phi)$.(Generally, it is assumed that
$V(\Phi^\ast\Phi)$ is a smooth function, for instance, a
polynomial~\cite{int}).

\par
Upto now we have considered the cases of charged scalar fields interacting
through a potential. It is also possible to consider interaction mediated
by a dynamical electromagnetic field $A_{\mu}$. The fields in the system
are now $\Phi$ (charged scalar fields) and $A_{\mu}$. If one applies an
external magnetic field $A_{ext}$ then the full Lagrangian is given by
\begin{equation}
{\cal L}= -\frac{1}{4} F^2 +(D_{\mu}\Phi)^\ast(D^\mu\Phi) -
m^2(\Phi^\ast\Phi) - V(\Phi^\ast,\Phi)
\end{equation}
 where $D_{\mu} =\partial_{\mu}-{\rm i} e A_{\mu}^{ext}-{\rm i} e A_{\mu}$ and
$F_{\mu\nu}=\partial_{\mu}A_{\nu} - \partial_{\mu}A_{\nu}$.

 The
argument given above can be modified as follows. The definition of $S_{0}$
changes slightly while $S_{int}$ remains the same.
  \begin{equation}
   {\rm S_{0}}=\int\int d\,^2x d\tau\left[-\frac{1}{4} F^2+(\partial_\mu-ie
A_\mu\Phi)^\ast(\partial^\mu+ie A^\mu\Phi) + m^2(\Phi^\ast\Phi)
                  +V(\Phi^\ast\Phi)\right],
\end{equation}
and
  \begin{equation}
  {\rm S_{int}}= \int\int d\,^2x d\tau\left[{-\rm i} e
(\partial_\mu\Phi)(A^\mu\Phi^\ast)
                  +{\rm i}e(A_\mu\Phi)(\partial^\mu\Phi^\ast)+e^2({\bf
A\cdot\bf A})(\Phi\Phi^\ast)\right].
\end{equation}
Again one can repeat the same argument by noting that $exp(-S_{0})$ is a
positive measure and the ratio $Z(B)/Z(0)$ as an expectation value
of $exp(-S_{int})$ to establish the diamagnetic inequality.

\section{Conclusions and Perspectives}
\par
 The response of a system to an electric field is completely different
 from its response to a magnetic field.
 The basic difference between the responses of a system on
 application of an electric field
 or a magnetic field lies in the Hamiltonian of the system.

 The Lagrangian of a system in the presence of an electric field
 can be written as
 \begin{equation}
 {\cal L}=(D_{0}\Phi)^\ast(D_{0}\Phi) -
(\nabla\Phi)^\ast(\nabla\Phi)-m^2(\Phi^\ast\Phi)- V(\Phi^\ast\Phi)
 \end{equation}
    where
 \begin{equation}
  D_{0} = \partial_{0}-{\rm i} e A_{0}
 \end{equation}
For statistical mechanics to make sense, the Hamiltonian $H$ must be
independent of time. Then it follows that

 \begin{equation}
{\cal H} =(\Pi^\ast)(\Pi)+(\nabla\Phi)^\ast(\nabla\Phi)+
m^2(\Phi^\ast\Phi)+V(\Phi^\ast\Phi)-{\rm
i}e\left[(\Pi^\ast)(A_{0}\Phi)-(\Pi)(A_{0}\Phi^\ast)\right]
 \end{equation}
 The electric field appears in the Hamiltonian
through the linear vector potential $A_{0}$ term. Now, from finite
temperature second order perturbation~\cite{lan1,lan2} theory, one can
show  easily that the free energy of the system always decreases with the
electric field. Hence, the dielectric susceptibility is always positive in
thermal equilibrium.
\par
But in the case of a magnetic field the Hamiltonian contains
both linear and quadratic terms in $A$. The net effect of an applied
magnetic field is not {\it a priori} clear. However, as our analysis makes
clear, for charged scalar field theories the net effect is always
diamagnetic.

In case of Spinless Bosons, there is no Zeeman term~\cite{zeem} coupled
with a magnetic field and hence the system consisting of Spinless Bosons
always has higher energy in a magnetic field than without the magnetic
field. It has been already pointed~\cite{sim,ss,bss,dav} out that there is
no corresponding theorem for Fermions.

\par
 Let us consider  some illustrative examples of Spinless Bose systems.
 One obvious example in the laboratory is the Cooper pair formation
in superconductors which shows  perfect diamagnetism (known as the
Meissner~\cite{sup} effect) below the critical temperature. Cooper pairs
also exist in Neutron Stars~\cite{neutr} where the magnetic field is very
high compared to any laboratory  field. Of course, the operators which
create and destroy Cooper pairs are not strictly Bose operators, so this
is only an analogy. It has been shown explicitly
by J. Daicic et. al.~\cite{mag} that the magnetised pair Bose systems
are relativistic superconductors. These systems are not covered by
previous analyses ~\cite{van,lan,sim,ss} which apply to non-relativistic
quantum mechanical systems. Pions would be a suitable candidate for
application of this theory with $\pi^+$ and $\pi^-$  regarded as the
particles and antiparticles. The choice is also motivated by the fact that
these pions are massive ($mc^2=139.5673$ Mev), obey Bose-Einstein statistics
and that they possess no spin. It is also well known that hard core
Bosons~\cite{bss} in any dimension on any lattice show a preference for
zero flux.
\par

\par

 Throughout this paper we have considered two spatial dimensions
both for the free and interacting cases, but the formalism easily generalizes
to higher dimensions. In  summary, we have shown exactly, that  charged
scalar fields at all  temperatures are diamagnetic.

\noindent
\newpage
\acknowledgements

It is a pleasure  to thank  Joseph  Samuel for suggesting this problem
and stimulating discussions;  Narendra Kumar, Rajaram Nityananda,
Chandrakant S Shukre, Diptiman Sen, Abhijit Khirsagar and Rajesh N.Parwani
for  useful discussions; Rangan Lahiri and Alexander Punnoose for critical
comments and P. Ramadurai for helping in plotting graphs.


\figure {The functions $G(B)$ and $H(B)$ have been plotted against $B$
for $L=3$ and $m=.1$. For simplicity we have drawn the figure for $B$
instead of $eB$.}

\figure {The function $K(B)=G(B)-H(B)$ has been plotted
against $B$. The values of $L$ and $m$ are the same as figure 1. Note that
the scale in this figure in $y$ direction is much expanded compared to
figure 1.}
\par
\figure {The full line is the curve  $f(\alpha)=\sqrt{.1+\alpha}$,
the dashed line is the tangent to the above curve at $\alpha=.5$. It is
straightforward to see that the area under the curve is less than the area
under the tangent. Also the area under the tangent is same as the area
of the rectangle defined by the dotted line. The area of this rectangle is
given by $f(1/2)\times 1=f(1/2)$. Hence, the positivity of $a(.1)$
is proved. This can be generalised to any positive value of $z_{l}$.}


\begin{references}
\bibitem{div1} This divergence of the susceptibility is reminiscent of the
fact that the free energy $F(B)\sim B^{3/2}$. This variation of the free
energy in this massless limit can also be understood from dimensional
argument.  Since we are working in natural units $\hbar=1$ and $c=1$, then
$[m]\sim[L]^{-1}$. Then the free energy density (i.e.per unit area )
varies as $[L]^{-3}$.  However, the dimension of $B$ is $[L]^{-2}$. So,
the massless limit restricts the free energy density variation with
magnetic field $B$ as $B^{3/2}$ only.
\bibitem {int} In field theory  one would also require that the
interaction $V(\Phi^\ast\Phi)$ be renormalizable. In (2+1)
dimensions  this would  restrict the interaction $(\Phi^\ast\Phi)^2$ or
 $(\Phi^\ast\Phi)^3$.  In condensed matter physics, no such restriction
apply since a natural cutoff exists.
\bibitem {zeem} For the case of relativistic Dirac electron, one finds a
Zeeman term $\sigma\cdot B$ in the Hamiltonian which ensures that the
energy will not necessarily increase with the magnetic field. So, in case
of fermions having spin the general tendency is to show
paramagnetism~\cite{para}.
Even  spinless fermions ~\cite{sim}, do not obey this diamagnetic
inequality. However , a corresponding Field Theory i.e. QED vacuum
in presence of an external homogeneous magnetic field is
diamagnetic~\cite{qed}.
\end{references}
\end{document}